\documentclass[sigconf, nonacm,dvipsnames]{acmart}

\usepackage{multirow}
\usepackage{tikz}
\usepackage{amsmath}
\usepackage{siunitx}
\usepackage[abbreviations]{foreign}
\usepackage{xspace}
\usepackage[inline]{enumitem}
\usepackage[algo2e,ruled,lined,boxed,commentsnumbered, noend, linesnumbered, procnumbered]{algorithm2e}
\SetKwComment{Cmt}{\color{gray}\sffamily\small // }{}
\definecolor{orange}{rgb}{1,0.5,0}
\definecolor{Green}{rgb}{0,0.5,0}
\definecolor{Blue}{rgb}{0,0,1}
\usepackage{svg}
\newcommand{\cmark}{\ding{51}}%
\newcommand{\xmark}{\ding{55}}%
\newcommand{\greencheck}{{\color{OliveGreen}\cmark}}
\newcommand{\redcross}{{\color{red}\xmark}}
 \usepackage{booktabs, makecell, pifont} 
\usepackage{acronym}
\acrodef{DL}{decentralized learning}
\acrodef{ML}{machine learning}
\acrodef{D-PSGD}{decentralized parallel stochastic gradient descent}
\acrodef{FL}{federated learning}
\acrodef{SGD}{stochastic gradient descent}
\acrodef{IID}{independent and identically distributed}
\acrodef{non-IID}{non independent and identically distributed}
\acrodef{RMSE}{root mean square error}
\acrodef{RMW}{random model walk}
\acrodef{GL}{gossip learning}
\acrodef{EL}{epidemic learning}
\acrodef{DWT}{discrete wavelet transform}
\acrodef{FFT}{fast Fourier transform}
\acrodef{MI}{mutual information}
\acrodef{DP}{differential privacy}
\acrodef{VN}{virtual node}
\acrodef{RN}{real node}
\acrodef{LDP}{local differential privacy}
\acrodef{PNDP}{pairwise network differential privacy}
\acrodef{PNLDP}{pairwise network local differential privacy}
\acrodef{GI}{gradient inversion}
\acrodef{CML}{collaborative machine learning}
\acrodef{TPR}{true positive rate}
\acrodef{FPR}{false positive rate}
\acrodef{LLM}{large language model}
\acrodef{RAG}{retrieval-augmented generation}
\acrodef{ANN}{approximate nearest neighbor}
\acrodef{LRU}{least-recently used}
\acrodef{FIFO}{first-in, first-out}
\acrodef{MMLU}{Massive Multitask Language Understanding}
\acrodef{NNS}{nearest neighbor search}
\acrodef{LSH}{locality-sensitive hashing}
\acrodef{TTFT}{Time To First Token}
\acrodef{HNSW}{hierarchical navigable small world} %
\newcommand{\sys}{\textsc{CatapultDB}\xspace}

\newcommand{\medragzipf}{\textsc{MedRAG-Zipf}\xspace}
\newcommand{\tripclick}{\textsc{TripClick}\xspace}
\newcommand{\pubmed}{\textsc{PubMed}\xspace}
\newcommand{\diskann}{\textsc{DiskANN}\xspace}
\newcommand{\freshdiskann}{\textsc{FreshDiskANN}\xspace}
\newcommand{\lshapg}{\textsc{LSH-APG}\xspace}
\newcommand{\proximity}{\textsc{Proximity}\xspace}
\newcommand{\vamana}{\textsc{Vamana}\xspace}
\newcommand{\freshvamana}{\textsc{FreshVamana}\xspace}
\newcommand{\filteredvamana}{\textsc{FilteredVamana}\xspace} %
\graphicspath{ {figures/} }
\usepackage{tikz}
\usepackage[eulergreek]{sansmath}
\usepackage{pgfplots}
\usepackage{pgfplotstable}
\usepackage{cleveref}
\usepackage{comment}
\crefname{assumption}{assumption}{assumptions}

\pgfplotsset{compat=newest}
\usepgfplotslibrary{external,units,colorbrewer,groupplots,fillbetween,statistics}
\usetikzlibrary{arrows.meta, backgrounds, fit, shapes.geometric, positioning}
\tikzsetexternalprefix{figures/}
\tikzset{external/mode=list and make}
\usetikzlibrary{patterns,shapes.misc}

\def\overleafhome{/tmp}
\newcommand{\inputplot}[2]{%
	\ifx\homepath\overleafhome%
	\IfBeginWith{#1}{plots}{\includegraphics{main-figure#2.pdf}}{#1}%
	\else%
	{\sffamily\scriptsize\input{#1}}
	\fi
}

\newcommand{\newgroupwidth}[2]%
{\expandafter\xdef\csname groupwidth#1\endcsname{#2}}

\newcounter{groupwidth}
\newsavebox{\groupwidthbox}
\makeatletter
{\edef\groupnumber{#1}%
	\stepcounter{groupwidth}%
	\@ifundefined{groupwidth\thegroupwidth}{\pgfmathsetlengthmacro{\mywidth}{\linewidth/\groupnumber}}%
	{\expandafter\let\expandafter\mywidth\csname groupwidth\thegroupwidth\endcsname}%
	\begin{lrbox}{\groupwidthbox}%
		\tikzset{/pgfplots/width={\mywidth}}%
		\ignorespaces}%
	{\end{lrbox}%
	\usebox\groupwidthbox
	\pgfmathsetlengthmacro{\mywidth}{\mywidth + (\linewidth - \wd\groupwidthbox)/\groupnumber}
	\immediate\write\@auxout{\string\newgroupwidth{\thegroupwidth}{\mywidth}}}
\makeatother
\usepackage{amsmath,amssymb,amsfonts,amsthm}
\usepackage{enumitem}
\usepackage{thm-restate}
\usepackage{bbm}
\theoremstyle{definition}

\theoremstyle{remark}

\allowdisplaybreaks

\makeatletter
\AtBeginDocument{%
	\def\ltx@label#1{\cref@label{#1}}%

	\def\label@in@display@noarg#1{\cref@old@label@in@display{#1}}%

	\def\label@in@mmeasure@noarg#1{%
		\begingroup
		\measuring@false
		\cref@old@label@in@display{#1}%
		\endgroup
	}%
}
\makeatother
 
\begin{document}
\title{Catapults to the Rescue: Accelerating Vector Search by Exploiting Query Locality}

\author{Sami Abuzakuk}
\orcid{0009-0003-6207-5905}
\affiliation{
	\institution{EPFL}
	\city{Lausanne}
	\country{Switzerland}
}
\email{sami.abuzakuk@epfl.ch}

\author{Anne-Marie Kermarrec}
\orcid{0000-0001-8187-724X}
\affiliation{
	\institution{EPFL}
	\city{Lausanne}
	\country{Switzerland}
}
\email{anne-marie.kermarrec@epfl.ch}

\author{Rafael Pires}
\orcid{0000-0002-7826-1599}
\affiliation{
	\institution{EPFL}
	\city{Lausanne}
	\country{Switzerland}
}
\email{rafael.pires@epfl.ch}

\author{Mathis Randl}
\orcid{0009-0003-6844-4695}
\affiliation{
	\institution{EPFL}
	\city{Lausanne}
	\country{Switzerland}
}
\email{mathis.randl@epfl.ch}
\authornote{Corresponding author}

\author{Martijn de Vos}
\orcid{0000-0003-4157-4847}
\affiliation{
	\institution{EPFL}
	\city{Lausanne}
	\country{Switzerland}
}
\email{martijn.devos@epfl.ch}

\begin{abstract}
Graph-based indexing is the dominant approach for approximate nearest neighbor search in vector databases, offering high recall with low latency across billions of vectors.
However, in such indices, the edge set of the proximity graph is only modified to reflect changes in the indexed data, never to adapt to the query workload.
This is wasteful: real-world query streams exhibit strong spatial and temporal locality, yet every query must re-traverse the same intermediate hops from fixed or random entry points.
We present \sys{}, a lightweight mechanism that, for the first time, dynamically determines where to begin the search in an ANN index on the fly, therefore exploiting query locality.
\sys{} injects shortcut edges called \emph{catapults} that connect query regions to frequently visited destination nodes.
Catapults are maintained as an additional layer on top of the graph, so the standard vector search algorithm remains unchanged: queries are simply routed to a better starting point when an appropriate catapult exists.
This transparent design preserves the full feature set of the underlying system, including filtered search, dynamic insertions, and disk-resident indices.
We implement \sys and evaluate it using four workloads with varying amounts of bias.
Our experiments show that \sys increases throughput by up to $2.51\times$ compared to \diskann at equivalent or better recall, matches the efficiency of \ac{LSH}-based approaches without sacrificing filtering or requiring index reconstruction, and adapts gracefully to workload shifts, unlike cache-based alternatives.
\end{abstract}

\maketitle

\section{Introduction}
\label{sec:intro}

\Acf{NNS} is a key operation in many domains, including recommendation systems~\cite{airen2022movie}, large-scale clustering~\cite{liu2007clustering}, similarity search~\cite{johnson2019billion}, and, more recently, \ac{RAG} pipelines for \ac{ML} inference~\cite{lewis2020retrieval}.
As vector databases scale to accommodate billions of high-dimensional data points, exact search becomes prohibitively expensive~\cite{jegou:2011:pq}.
This motivates \ac{ANN} search: rather than guaranteeing exact results, these methods return points that are \emph{close enough} to the query, trading a small, bounded loss in accuracy for significant reductions in query latency~\cite{indyk1998approximate}.
The dominant paradigm for achieving high throughput and low latency in \ac{ANN} search is to build a graph-based index~\cite{malkov:2014:nsw,malkov:2020:hnsw}.
This involves the construction of a proximity graph in which vectors are nodes and edges connect approximate nearest neighbors.
Search is then performed via a greedy traversal from a designated entry point, navigating the graph until converging on the local neighborhood closest to the query vector.

\begin{table*}[t]
    \centering
    \caption{Feature comparison of vector search approaches.}
    \label{tab:comparison}
    \begin{tabular}{lcccc}
        \toprule
        \textbf{System}
        & \makecell{\textbf{Accelerated}\\\textbf{search (LSH)}}
        & \textbf{Supports insertions}
        & \textbf{Supports filtering}
        & \makecell{\textbf{Scales to}\\\textbf{persistent memory}}\\
        \midrule
        LSH-APG~\cite{zhao:2023:lshapg}
        & \greencheck & \greencheck & \redcross & \redcross \\
        Proximity~\cite{bergman:2025:proximity}
        & \greencheck & \redcross & \redcross & \greencheck \\
        DiskANN~\cite{subramanya:2019:diskann}
        & \redcross & \greencheck & \greencheck & \greencheck \\
        \bottomrule
        \textbf{\sys~(ours)}
        & \greencheck & \greencheck & \greencheck & \greencheck \\
        \bottomrule
    \end{tabular}
\end{table*}

Despite these advances, graph-based \ac{ANN} methods share an implicit constraint: the edge set of the proximity graph is modified only when the \emph{set of nodes} changes, \eg, when new documents are added to the database~\cite{subramanya:2019:diskann,malkov:2020:hnsw,singh:2021:freshdiskann}.
The vast majority of vector database implementations rely on an edge structure that is determined at construction time and, once built, remains static with respect to the query workload.
This is sub-optimal since real-world workloads exhibit strong temporal and spatial locality: queries cluster in certain regions of the vector space, and the same exploration paths are redundantly traversed by successive, similar queries~\cite{petersen:2016:power,silverstein:1999:analysis,rekabsaz:2021:tripclick,goccer2026qvcache}.
Yet, every query must begin from an entry point that is oblivious to the workload and re-traverse redundant intermediate hops, incurring unnecessary latency and compute overhead per lookup.

Prior work has attempted to address this inefficiency from two angles.
The first angle is to select better entry points into the graph, reducing the length of the traversal path from the entry point to the query's neighborhood.
For example, \lshapg~\cite{zhao:2023:lshapg} uses \acf{LSH}~\cite{gionis:1999:lsh} to determine better entry points into the graph, reducing the length of the traversal path.
However, \lshapg's shortcuts are computed statically at index-construction time from the data distribution alone, with no awareness of the query workload: its entry-point index must be built from scratch alongside the proximity graph, so it cannot be layered on top of an existing index.
This also means \lshapg cannot exploit runtime information such as attribute-based filter predicates, which are common in production and only known at query time~\cite{gollapudi:2023:filtereddiskann}.
Finally, because the hash functions in \lshapg are calibrated to the range of possible distances in both the query and indexed datasets at construction time, any out-of-distribution insertion degrades entry-point quality and ultimately requires full index reconstruction, which is a prohibitive cost in settings with frequent data updates.

The second angle is to exploit query locality more directly by caching results from previous queries: if an incoming query is sufficiently similar to a past one, the cached result is returned without a lookup~\cite{bergman:2025:proximity,goccer2026qvcache}.
For example, \proximity~\cite{bergman:2025:proximity} introduces an approximate cache that caches queries and results.
If an incoming query is within some distance threshold of a previously seen query, the cached result is returned directly, bypassing the database entirely.
While \proximity is effective at reducing query lookup time for workloads with temporal locality, it is not aware of dynamic insertions into the underlying vector database, and its sensitivity to a distance threshold parameter means that when the workload distribution shifts, cached entries become stale and results degrade. %
Table~\ref{tab:comparison} summarizes these discussed trade-offs.
An ideal solution would combine the entry-point quality of \lshapg with the workload-awareness of the \proximity cache. %

This paper introduces \sys, a novel approach that, for the first time, dynamically modifies the edge set of a proximity graph  to exploit temporal and spatial locality of query workloads.
\sys adapts to recurring patterns in search trajectories: when queries in a region of the vector space (identified using LSH) repeatedly traverse from an entry point toward the same
destination node, \sys injects a direct shortcut edge, which we call a \emph{catapult}, connecting the two.
Future queries in the same region are \emph{catapulted} directly to the relevant neighborhood, skipping the redundant intermediate hops entirely.

A key design principle of \sys{} is transparency: it operates as a lightweight layer on top of any index that accepts a \emph{hint} for where to begin the search, such as the entry node in \diskann~\cite{subramanya:2019:diskann} or HNSW~\cite{malkov:2020:hnsw}.
\ac{NNS} still proceeds via the normal greedy traversal algorithm, with queries simply gaining immediate access to a better area of the graph when historical locality information is available.
Because catapults are edges in the graph itself rather than entries in an external cache, \sys does not compromise on any of capabilities that the underlying index provides, including dynamic insertions and deletions, attribute-based filtered search~\cite{gollapudi:2023:filtereddiskann}, and scaling to billion-point datasets on commodity hardware (which involves storing a part of the index on persistent storage, as opposed to relying entirely on the limited system memory) ~\cite{subramanya:2019:diskann,singh:2021:freshdiskann,gollapudi:2023:filtereddiskann}.

We implement \sys as a layer on top of any graph-based \ac{ANN} index that exposes a configurable entry point; in our evaluation, we use \diskann, a popular and widely-deployed vector database.
We evaluate \sys using four workloads with varying amounts of locality.
Our main experimental findings are:
\begin{enumerate}
    \item \sys increases throughput by $2.51\times$ compared to vanilla \diskann and by $2.97\times$ compared to \lshapg at equivalent or better recall, by exploiting workload locality to significantly reduce graph traversal paths per query.
    Even in the worst case where there is no locality in queries, \sys generally does not compromise throughput compared to baselines.
    \item \sys naturally supports filtered queries and dynamic insertions without requiring index reconstruction.
    Compared to \diskann, \sys achieves up to $1.38\times$ higher throughput and obtains up to $11\%$ higher recall.
\end{enumerate}

In summary, our key contributions are:
\begin{enumerate}
    \item We identify the opportunity for online, query-driven edge addition in graph-based \ac{ANN} indices and introduce the notion of \emph{catapults}, \ie, shortcut edges that are injected into the graph based on observed search trajectories.
    To the best of our knowledge, \sys is the first system to dynamically modify the edge set of a proximity graph to exploit the locality of the query workload, rather than solely for reflecting changes to the indexed data (\Cref{sec:design}).
    \item We implement \sys as an index-agnostic layer that transparently augments any vector index accepting a search hint, preserving the full feature set of the underlying system, including filtered search and dynamic insertions (\Cref{sec:exp_setup}).
    \item We conduct extensive experiments using four workloads with varying degrees of locality (\Cref{sec:experiments}).
    Our results demonstrate that \sys significantly improves query throughput compared to \diskann and \lshapg, natively supports filtered queries (unlike \lshapg), and adapts robustly to shifting workloads, unlike cache-based approaches.
\end{enumerate}

\section{Background and motivation}
\label{sec:background}
We first introduce \acf{NNS} and graph-based \acf{ANN} search on which \sys is built and then describe \acf{LSH}, which \sys uses to identify query regions and route queries to better entry points in the index.
Finally, we motivate the key insight behind \sys by characterizing locality properties observed in real-world query workloads.

\subsection{Vector search}
\label{sec:vectorsearch}

\subsubsection{\Acf{NNS}}
\Ac{NNS} is the problem of finding the most similar vectors to a given query in a large collection~\cite{friedman1977algorithm}.
Formally, let $\mathcal{D} \subset \mathbb{R}^d$ be a dataset of $n$ vectors with dimensionality $d$.
Given a query vector $q \in \mathbb{R}^d$ and a positive integer $k$, a $k$-nearest neighbor ($k$-NN) query returns the $k$ vectors in $\mathcal{D}$ closest to $q$ under a given distance function (\eg Euclidean distance).  Exact $k$-NN search requires comparing $q$ to every vector in $\mathcal{D}$, which is prohibitively expensive when $n$ and $d$ are large.

\subsubsection{\Acf{ANN} search}
Since exact $k$-NN search does not scale to large datasets, \acf{ANN} search relaxes the exactness requirement in exchange
for significantly lower query latency~\cite{indyk1998approximate}.
A $(c, k)$-ANN query returns $k$ vectors $o_1, \ldots, o_k$ such that $\|q, o_i\| \leq c \cdot \|q, o_i^*\|$, where $o_i^*$ is the $i$-th true nearest neighbor and $c > 1$ is an approximation ratio.
In practice, ANN search methods are evaluated by \emph{recall}, \ie, the fraction of true $k$-NN that appear in the result,
and \emph{query latency}.
The three main families of \ac{ANN} search methods are tree-based (\eg KD-trees), quantization-based (\eg Product
Quantization~\cite{jegou:2011:pq}), and graph-based methods.
Graph-based \ac{ANN} search methods have consistently shown the best recall-vs-latency trade-offs in recent benchmarks and are widely used in production settings~\cite{wang:2021:survey,li:2020:survey}.
Therefore, they form the foundation of our work.

\subsubsection{Graph-based \ac{ANN} search}

Graph-based \ac{ANN} search methods build a proximity graph $G = (V, E)$ where each node corresponds to a vector in $\mathcal{D}$ and edges connect approximate neighbors.
Search proceeds by greedy traversal: starting from one or more entry points, the algorithm repeatedly moves to the unvisited neighbor closest to the query, expanding a candidate set until convergence.
This beam search procedure on graph-based indices is shown in \Cref{algo:search}.
A key parameter that impacts performance is the set of starting points $sp$: their quality directly determines how many hops
are needed to reach the query's neighborhood.
This method requires the graph to have a structure that enables this greedy search to find reasonable results, a property known as \emph{navigability}~\cite{diwan2024navigable}.

\begin{figure}
	\includegraphics{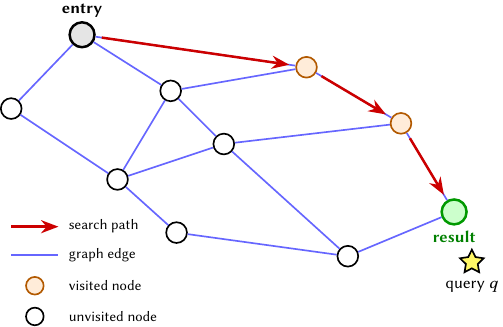}
	\caption{Standard beam search for $k=1$ neighbor. The search path is selected greedily, by hopping to the neighbor that minimizes the distance to the query.}
    \label{fig:beam_search}
\end{figure}

\begin{algorithm2e}[t]
	\caption{Graph-based beam search (adapted from DiskANN~\cite{subramanya:2019:diskann})}
	\label{algo:search}
	\DontPrintSemicolon
	\SetKwProg{Fn}{Procedure}{:}{}
	\SetKwInOut{Parameter}{Input}
	\SetKwInOut{Output}{Output}
	\Parameter{\quad query vector $q$, neighbor retrieval count $k$, initial \\\quad starting points $sp$}
	\Fn{\textsc{lookup}($q$, $k$, $sp$)}{
		$C = sp$ \Cmt*[r]{set of candidates}
		$V = \{\}$ \Cmt*[r]{set of visited nodes}
		\While{$C \setminus V \neq \{\}$} {
			$currentCandidate = x \in C \setminus V \text{ closest to } q$\\
   			$V = V \cup \{currentCandidate\}$\\
			$C = C \cup \textsc{neighbors}(currentCandidate)$\\
			$C = \textsc{trim}(C, k, q)$ \Cmt*[r]{keeps up to $k$ closest to $q$}
		}
		\Return $C$
	}
\end{algorithm2e}

\subsubsection{Filtered \ac{ANN}}
\label{sec:filtered_ann}
Many real-world applications attach metadata labels to vectors and require that query results satisfy a predicate on those labels, \eg ``return the $k$ nearest neighbors of $q$ among all vectors matching the predicate $f$'', where $f$ is a boolean function of the tags associated with the documents~\cite{gollapudi:2023:filtereddiskann}.
Formally, a \emph{filtered} $(c,k)$-ANN query returns $k$ vectors $o_1, \ldots, o_k \in \mathcal{D}_f = \{x \in \mathcal{D} : f \in F_x\}$ such that $\|q, o_i\| \leq c \cdot \|q, o_i^*\|$, where $o_i^*$ is the $i$-th true nearest neighbor within $\mathcal{D}_f$.
A naive approach post-filters the results of an unfiltered query, but this has no guarantee of correctness for low-specificity labels (rare filters) because the unfiltered search may return zero points matching the predicate.
\textsc{FilteredVamana}~\cite{gollapudi:2023:filtereddiskann} addresses this by incorporating label information directly into graph construction and search: each query starts from a dedicated per-label entry point, and the pruning rule is extended to preserve filter-subgraph navigability.
This ensures that the greedy traversal stays within the relevant subset of the graph, maintaining high recall even for filters covering less than 1\% of the dataset.
Major production vector databases, including Pinecone~\cite{pinecone}, Weaviate~\cite{weaviate}, and Milvus~\cite{wang2021milvus}, expose filtered search as a first-class feature.
Thus, supporting filtered search is an important requirement for any contemporary vector database. %

\subsection{\Acf{LSH}}
\label{sec:lsh}
One way to speed up graph-based \ac{ANN} is to use \acf{LSH} to identify a better entry point into the graph, reducing the number of hops needed to reach the query's neighborhood.
\ac{LSH}~\cite{gionis:1999:lsh} is a family of hash functions designed so that nearby points in the original space are more likely to collide (share the same hash value) than distant points.
Formally, a family $\mathcal{H} = \{h : \mathbb{R}^d \to \mathbb{R}\}$ is $(r, cr, p_1, p_2)$-locality-sensitive if, for any two points $o_1, o_2$: (i) $\|o_1, o_2\| \leq r$ implies
$\Pr[h(o_1) = h(o_2)] \geq p_1$, and (ii) $\|o_1, o_2\| > cr$ implies $\Pr[h(o_1) = h(o_2)] \leq p_2$, with $p_1 > p_2$.

In this work, we use \emph{random hyperplane} \ac{LSH} to partition the query space into regions (see \Cref{sec:design}).
In this \ac{LSH} instantiation, each hash function is defined by a random vector $\vec{r} \in \mathbb{R}^d$ drawn from the standard normal distribution.
This vector represents a normal to a $d$-dimensional hyperplane passing through the origin.
A point $o$ is mapped to a single bit: $h(o) = \mathbf{1}[\vec{r} \cdot \vec{o} \geq 0]$, representing the side of the hyperplane in which $o$ is located (a hash value of $1$ designating the same side as the normal $r$).
By concatenating $L$ such bits, one obtains an $L$-bit hash code that partitions the space
into $2^L$ buckets.
We note that this variant is also used by \proximity~\cite{bergman:2025:proximity} to index its cache.

\subsection{Temporal and spatial locality in real-world query workloads}
\label{sec:querybias}

\begin{figure}
	\includegraphics{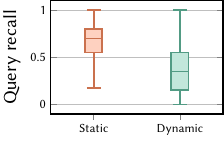}
	\caption{The recall distribution of queries of \proximity when dealing with a static vector database (left) vs. a dynamic one (right).}
    \label{fig:killprox}

\end{figure}

Real-world query workloads exhibit a high degree of temporal and spatial locality.
At the lexical level, it is well-known that query frequency in large-scale search logs follows a heavy-tailed, approximately power-law (Zipfian) distribution: a small number of queries are submitted extremely frequently, while the vast majority are rare~\cite{silverstein:1999:analysis,petersen:2016:power}.
This structural skew has been consistently observed across diverse domains and scales, from general-purpose web search logs~\cite{silverstein:1999:analysis} to health-domain retrieval platforms.
For example, in \textsc{TripClick}~\cite{rekabsaz:2021:tripclick}, a dataset of 5.2 million user interactions collected from a medical search engine over seven years, the query frequency histogram follows an exponential trend: a handful of queries such as \emph{``asthma pregnancy''} account for a disproportionate share of traffic, while the long tail consists of queries that are issued only a few times.

Since semantically similar queries map to nearby vectors under any reasonable embedding model, lexical skew in query logs translates directly into spatial skew in the vector space.
When user queries are encoded as dense vectors and submitted to a vector database, the resulting query workload concentrates in a small number of regions of the high-dimensional space.
Multiple studies on industrial vector search workloads confirm that access patterns over an index are highly non-uniform~\cite{tian2024scalable,mohoney:2024:incremental,goccer2026qvcache}.
A study on Knowledge Graph entity search workloads finds that only about 15\% of the Inverted File (IVF) indexes were accessed over an entire day of traffic~\cite{mohoney:2024:incremental}.
This access skew has practical consequences for system design: \textsc{QVCache}~\cite{goccer2026qvcache} demonstrates that a megabyte-scale cache suffices to serve a significant fraction of queries against a billion-point index, precisely because the active working set is small relative to the full dataset.

Beyond spatial concentration, workloads also exhibit strong \emph{temporal} locality: successive queries tend to cluster together in the embedding space, with bursts of semantically similar queries arriving in short time windows~\cite{frieder2024caching}.
This behavior, for example, arises naturally in conversational applications, where users ask similar,  related questions within a single session, or in e-commerce, where multiple users concurrently search for the same trending product (\eg, air conditioners in summer)~\cite{chien2005semantic}.
Yet because successive queries are never exact duplicates, each is treated as an independent lookup, leaving the spatial and temporal structure of the workload entirely unexploited.

Some systems, such as \proximity, exploit the locality of queries by deploying an approximate cache as a middleware between the vector database and the application layer.
It observes traffic between the two and intercepts queries that are sufficiently similar to previous ones stored in cache.
It proceeds to return immediately the stored corresponding documents.
Such systems do not tolerate insertions and deletions in the underlying database: they either delete their state at every database update, therefore losing track of all accumulated history, or they serve out-of-date results.
This makes them unsuitable for any vector database with ongoing data ingestion. %
We experimentally demonstrate this by first populating the vector database with embeddings generated by the \pubmed dataset and then sending the \medragzipf query workload (further described in \Cref{sec:exp_setup}) to \proximity.
In one experiment run, the vectors stored in the database remain static whereas in the other, new vectors are dynamically inserted (a batch of \num{5000} vectors for every 50 queries), causing \proximity to serve outdated results.
\Cref{fig:killprox} shows the recall for both the static and dynamic run.
The recall degrades dramatically when the content in the vector database changes, with the median recall dropping from 70\% with a static vector database to 35\% with a dynamic one.

\section{Design of \sys}
\label{sec:design}

Motivated by the locality properties of real-world query workloads, we design \sys around four goals:
\begin{enumerate*}[label=\emph{(\roman*)}]
\item \emph{workload-awareness}, \ie, exploiting spatial and temporal locality in the query stream to reduce redundant graph traversal; 
\item \emph{support for dynamic updates}, remaining efficient as vectors are inserted or deleted without requiring index reconstruction;
\item \emph{support for filtered queries}; and
\item \emph{index-agnosticism}, operating as a lightweight layer on top of any existing graph-based index that exposes a configurable entry point, requiring no changes to the underlying search algorithm.
\end{enumerate*}

We first explain the high-level idea and insight behind \sys in \Cref{sec:nutshell} and then provide a detailed explanation of the components in \sys in the following subsections.

\begin{figure}[t]
	\includegraphics{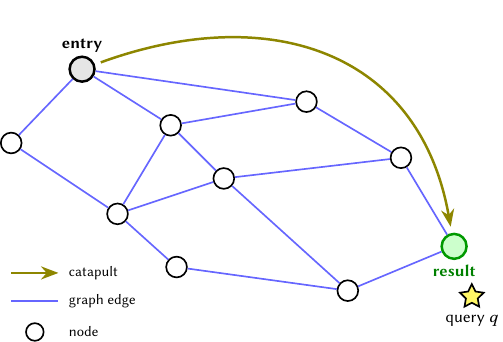}
	\caption{A catapult-accelerated beam search for $k=1$ neighbor. Instead of traversing multiple intermediate hops from the entry point, the query is launched directly into its target neighborhood via a catapult edge, significantly reducing the search path length.}
    \label{fig:catapult}
\end{figure}

\subsection{\sys in a nutshell}
\label{sec:nutshell}
Because real-world queries are spatially concentrated and temporally correlated, information gathered from recent queries in a region of the vector space is a strong predictor of the traversal paths that future queries in that region will follow.
This is the core insight behind \sys: rather than discarding the result of each traversal after the query completes, \sys records it and uses it to shortcut the starting point of future queries in the same region, directly exploiting the workload locality that static indices ignore.
We call these shortcuts \emph{catapults}: directed edges that connect a query region (identified by its \ac{LSH} hash code) to a better entry point in the proximity graph.
This process is visualized in \Cref{fig:catapult}.

We design \sys as a lightweight layer that dynamically adapts the entry points of an underlying graph-based ANN index to the observed query workload.
Catapults are maintained as an auxiliary layer on top of the index and are passed as additional starting points to a standard \ac{ANN} procedure (\eg Algorithm~\ref{algo:search}).
The underlying search algorithm remains entirely unchanged. %

\subsection{The catapult mechanism}
\label{sec:catapult-management}
We next provide a detailed description of the catapult subsystem.

\begin{algorithm2e}[t]
	\caption{Catapulting layer on top of existing search}
	\label{algo:catapulted-search}
	\DontPrintSemicolon
	\SetKwProg{Fn}{Procedure}{:}{}
	\SetKwInOut{Parameter}{Input}
	\SetKwInOut{Output}{Output}
	\Parameter{\quad query vector $q$, neighbor retrieval count $k$, \\\quad graph medoid $m$, initially empty $buckets$, \\\quad bucket capacity $b$}

    \Fn{\textsc{catapulted\_lookup}($q$, $k$, $m$, $lsh$, $buckets$)}{
		$hash = \textsc{lsh}(q)$ \\
		$catapults = buckets[hash]$ \\
		$sp = catapults \cup \{m\}$ \Cmt*[r]{Starting points}
        $ann = \textsc{Lookup}(q, k, sp)$ \Cmt*[r]{See \Cref{algo:search}}
        $catapults\text{.\textsc{append}}(ann\text{.\textsc{best}()})$\\
        \If{$|catapults| > b$}{$catapults$.\textsc{lru\_evict}()}
        \Return ann
	}

\end{algorithm2e}

\paragraph{Catapulted lookups.}
\Cref{algo:catapulted-search} shows the \textsc{catapulted\_lookup} procedure to perform a \ac{ANN} search with \sys.
This algorithm calls the \textsc{lookup} procedure in \Cref{algo:search}, which is exposed by the underlying database, but creates and manages catapults to speedup vector search.
The \textsc{catapulted\_lookup} procedure takes as input the query vector $ q $, the number of neighbors $k$ we wish to retrieve, the graph medoid $m$, and a list of buckets that contain catapults.
The graph medoid, which is the vector closest to all others on average, is included to account for cases where the catapult set is empty or irrelevant to the query.
\sys uses \ac{LSH} to organize search entry points into buckets, which is discussed in more detail later in this section.

Given an incoming query $q$, its \ac{LSH} hash code is computed, and the corresponding bucket is retrieved (lines~2--3).
The contents of the bucket, together with the \emph{medoid} $m$ of the dataset, form the set of starting points $sp$ that are passed to the underlying beam search (line~4).
After the \textsc{lookup} procedure completes and returns a list $ann$ of up to $k$ vectors considered close to $q$, the best result in $ann$ (\ie the node in the proximity graph closest to $q$ among the returned neighbors) is appended to the catapults in the bucket (line~6).
This node serves as a catapult destination for future queries that map to the same \ac{LSH} region, effectively creating a new catapult.
Instead of initiating traversal from the medoid $m$ and incurring multiple hops across the proximity graph, subsequent queries that are similar are catapulted directly into the relevant neighborhood, thus exploiting spatial locality.
We note that injecting catapults as entry points is equivalent to adding an edge from the medoid to the catapult’s landing node upon an \ac{LSH} match.
Our approach, however, allows us to decouple the catapult mechanism and the beam search (\Cref{algo:search}), since their interaction is limited to the starting-point interface.

Note that in the case where the query $q$ has never been seen by \sys but maps to a non-empty \ac{LSH} bucket, the entry points provided to the internal search method may still turn out to be useful by serendipity:
the catapult landing points stored in that bucket were recorded from past queries in the same \ac{LSH} region and are therefore more likely to lie near $q$'s neighborhood than the graph medoid.
In particular, the graph medoid has been chosen to minimize the number of hops required to reach any point by greedy navigation, with no regard for the query distribution, which \sys leverages instead.

\paragraph{LSH-based region identification.}
Since graph-based \ac{ANN} indices typically initiate each query from a single fixed entry point in the proximity graph, naively adding catapult edges from that node to all targeted regions of the graph would overload it with a large number of outgoing edges.
All of these edges would then need to be evaluated before performing the first hop, increasing the initial search overhead by a large margin.

To significantly reduce the computational cost of processing a query,
we use \ac{LSH} to compute the similarity between the current query and the previous query that generated the catapult.
In practice, \sys partitions the query space into regions using random hyperplane \ac{LSH}.
Given an incoming query vector $q \in \mathbb{R}^d$ and $L$ randomly generated hyperplanes with normal vectors $r_1, \ldots, r_L$, the \ac{LSH} hash code of $q$ is $(q \cdot r_1 \geq 0,\ \ldots,\ q \cdot r_L \geq 0)$ as described in ~\cref{sec:lsh}.
Queries that share a hash code are considered to belong to the same region and are handled by the same catapult bucket.
As random hyperplane \ac{LSH} is scale-invariant, no dataset-specific calibration is required, in contrast to the $p$-stable \ac{LSH} variant used by \lshapg.

\paragraph{Catapult eviction.}
Each bucket is assigned a fixed capacity, defined as a global, pre-determined constant $b$ (see \Cref{sec:parameters}).
When the bucket is full, and a new catapult is to be inserted, an existing entry is evicted (lines~7--8 of \Cref{algo:catapulted-search}) using the \ac{LRU} eviction policy.
This eviction is key for \sys to adapt to workload shifts: evicting the least-recently used entry ensures that destinations associated with older, possibly stale query patterns are replaced by those relevant to the current workload.
This design also provides adaptivity to document insertions in the underlying graph: as new vectors are added to the index, they may constitute better catapult destinations than those currently stored in a bucket.
Because recently issued queries will naturally land on these new nodes during traversal, the corresponding bucket entries are refreshed over time, and stale destinations pointing to previously optimal but now superseded nodes are gradually evicted.
No explicit notification of index updates is required; the eviction mechanism adapts passively through the normal query stream.

\paragraph{Proximity graph creation.}
\sys operates as a layer on top of an existing proximity graph and does not require constructing the graph from scratch.
In the evaluation of \sys (see \Cref{sec:experiments}), we use \vamana, the graph construction algorithm used by \diskann.
\vamana offers strong empirical performance and is likely already available in deployments where \sys would be adopted.

\paragraph{Competitive recall.}
Graph-based \ac{ANN} indices are constructed to be navigable, \ie, greedily traversable, meaning that the choice of entry point affects only the number of hops required to reach the query's neighborhood, not whether that neighborhood is ultimately found.
Recall is therefore robust to the quality of the starting points: a poor catapult merely means that the traversal starts farther from the target and requires more hops, potentially degrading throughput but never correctness.
This property is reinforced by the inclusion of the graph medoid in the starting point set (line~4 of Algorithm~\ref{algo:catapulted-search}).
Even when no catapult has been recorded for a given \ac{LSH} region, the medoid provides the same baseline as an unmodified \diskann search.
Catapults are designed to offer a non-negative benefit: when informative, they enhance throughput; when not, they ideally fall back to baseline behavior.
In practice, a small throughput regression can occur in the rare case when there is no locality in the query workload (see \Cref{sec:exp_unbiased_workload}).

\paragraph{Negligible storage cost.}
Catapult destinations are stored as node identifiers in the proximity graph and incur negligible memory overhead: a bucket of size $b$ only requires storing $b$ 4-byte indices per \ac{LSH} bucket, for a total of $b \cdot 2^L$ integers
across all $2^L$ buckets.
For the typical values of $b = 40$ and $L = 8$ used across this paper, this amounts to bounding the memory requirements of \sys to \qty{40}{\kibi\byte} of extra edge data over the requirements of the underlying beam search algorithm.
In practice, this easily falls within the memory budget of conventional deployments.
In comparison, graph vector indices frequently reach the tens or hundreds of gigabytes, mostly due to the storage of full-precision and compressed vectors~\cite{pan:2024:survey}.

\paragraph{Integration with \diskann}
\diskann~\cite{subramanya:2019:diskann} stores the proximity graph on SSD while maintaining only compressed vector representations in memory, enabling it to scale to billion-point datasets on commodity hardware. Since the catapult subsystem only provides entry points and delegates the search to the underlying beam search implementation, \sys naturally inherits this scalability from \diskann.

Because the number of hops per query determines the number of SSD reads, reducing the hop count translates directly into lower I/O latency.
Catapults further amplify this effect: by launching the search closer to the target neighborhood, \sys reduces the hop count and therefore the number of disk reads, reinforcing the latency savings of  \diskann.

\subsection{\sys parameters}
\label{sec:parameters}

\sys introduces two system parameters: the number of \ac{LSH} hyperplanes $L$ and the bucket capacity $b$.
We discuss the impact of these two parameters on the performance of \sys.

\paragraph{Number of \ac{LSH} hyperplanes ($L$).}
$L$ controls the granularity with which \sys partitions the query space.
Each additional hyperplane doubles the number of buckets, yielding $2^L$ regions in total.
A larger value of $L$ produces finer-grained regions, reducing the chance that two dissimilar queries share a bucket and are served the same catapult destinations.
However, increasing $L$ also increases the maximum memory consumption linearly in $2^L$ because each additional hyperplane doubles the number of buckets, and risks making individual buckets so specific that they are rarely reused, undermining the exploitation of query locality that motivates \sys.
A smaller value of $L$ produces coarser regions with higher reuse but potentially noisier catapult destinations.
In general, $L$ should be tuned so that the expected number of queries per bucket is large enough to populate catapults quickly, while remaining small enough that buckets correspond to meaningfully coherent regions of the query space.
We empirically find that $L=8$ offers the highest end-to-end throughput across our experiments (see \Cref{sec:exp_hyperparams}).

\paragraph{Bucket capacity ($b$).}
$b$ denotes the maximum number of catapult destinations retained per \ac{LSH} bucket and governs the tradeoff between adaptivity and stability.
A small value of $b$ allows \sys to react quickly to workload shifts, \ie, stale destinations are evicted promptly as new queries arrive, but may discard useful catapults when queries within the same region are diverse.
A large value of $b$ accumulates more historical information in a bucket and achieves higher hit rates under stable workloads, at the cost of evaluating more candidate starting points per query.
We empirically find that $b=40$ offers the highest end-to-end throughput across our experiments (see \Cref{sec:exp_hyperparams}).

\paragraph{Underlying graph properties.}
Beyond \sys's own parameters, the properties of the underlying graph index influence how much benefit catapults can deliver.
We next discuss important parameters in a graph created by \vamana since we primarily evaluate the performance of \sys with \diskann.
The underlying \vamana graph exposes three key parameters: the maximum out-degree $R$, the pruning parameter $\alpha$, and the dataset size $N$.
These three parameters are properties of the \vamana graph and are not set by \sys itself, but they influence how much benefit catapults can deliver.
The maximum out-degree $R$ controls the number of neighbors evaluated at each hop: a higher-degree graph explores more of the neighborhood per step, at the cost of spending more time per step to evaluate its larger set of neighbors.
$\alpha > 1$ is the \vamana pruning parameter that trades a modest increase in degree for a substantially smaller graph diameter, reducing the number of hops required to reach any query's neighborhood from the medoid.
A graph built with a large value of $\alpha$ already has short traversal paths, compressing the absolute savings that catapults can achieve; conversely, a lower-$\alpha$ graph with longer baseline traversals offers more headroom for catapult-driven reduction.
Finally, the dataset size $N$ affects the coarseness of \ac{LSH} bucket coverage relative to the graph: as the dataset grows, the $2^L$ buckets span proportionally larger regions of the vector space, which may reduce the precision of catapult destinations and suggest increasing $L$ accordingly.

\begin{figure}[t]
	\includegraphics{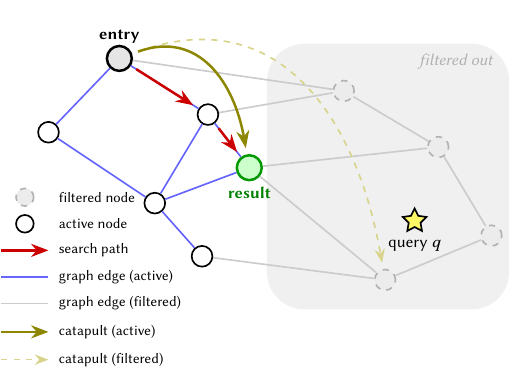}
	\caption{Processing a filtered query in \sys. A filter is associated to the query at runtime, disabling some parts of the proximity graph. Catapults can be conditionally disabled.}
    \label{fig:filtered_search}
\end{figure}

\subsection{Support for filtering}
\label{sec:filtering}

As discussed in \Cref{sec:filtered_ann}, many vector search workloads associate metadata labels with indexed documents and require query results to satisfy a predicate over those labels. %
\filteredvamana~\cite{gollapudi:2023:filtereddiskann} addresses this by incorporating label information into graph construction and search: each query is assigned a per-label entry point, and the greedy traversal is constrained to nodes that satisfy the active filter.
\sys composes naturally with this mechanism.

When a filtered query arrives, its \ac{LSH} hash code is computed from the query vector as usual.
However, the catapult destinations stored in the corresponding bucket are also checked against the active filter predicate before being included in the starting point set.
A destination node that does not satisfy the predicate is excluded from $sp$, and the search falls back to the per-label entry point provided by \filteredvamana for that starting position.
We also visualize this process in \Cref{fig:filtered_search}.
This ensures that the greedy traversal remains within the filter-defined subgraph, preserving the correctness guarantees of \filteredvamana even when catapults are active.
Note that such a system cannot be instantiated at index-construction time. The space of possible query filters is exponentially larger than the set of document tags, making it intractable to pre-compute and store meaningful entry points for every potential filter combination.

Upon completion of a filtered query, the highest-ranked neighbor among the returned results, necessarily satisfying the filter predicate, is inserted into the bucket and tagged with the active filter.
Future queries with the same \ac{LSH} hash code and filter predicate are thus catapulted directly into the appropriate filtered neighborhood, amplifying the latency reductions achieved by both \sys and \filteredvamana.

\subsection{Multithreading and concurrency}
\label{sec:concurrency}

Production vector databases process queries concurrently, dispatching each incoming request to a separate CPU core.
All CPU cores share access to the same in-memory graph instance. %
\sys inherits this model: independent queries execute in parallel, each running the \textsc{Catapulted\_Lookup} procedure in \Cref{algo:catapulted-search} on its own thread without any coordination at the level of the underlying vector database.

Concurrency introduces a challenge that is unique to \sys among the systems compared in \Cref{tab:comparison}.
\diskann and \lshapg treat their index structures as read-only at query time, while \proximity is single-threaded.
Catapults, in contrast, are updated dynamically during query execution.
\sys must therefore synchronize accesses to the catapult data structure to prevent undefined behavior under multithreaded execution.
Beyond correctness, liveness should also be considered.
As discussed in \Cref{sec:background}, queries exhibit strong locality.
More specifically, bursts of spatially similar queries may arrive within a short time window.
To exploit this property, a completed query must publish its catapult in a way that makes it immediately visible to concurrently starting queries on other cores, thereby allowing the entire burst to benefit from the shortcut.

We satisfy both requirements using a reader-writer lock associated with each bucket. At the start of \textsc{Catapulted\_Lookup}, the thread acquires the corresponding bucket lock in read mode (line~3), allowing concurrent queries to read the same bucket in parallel, while preventing concurrent writes. It then reads the content of the bucket and releases the lock. When the underlying search completes, and a new catapult destination is ready to be published (lines~6-8), the thread re-acquires the same bucket lock in exclusive write mode.  This may require waiting for potential ongoing reading or writing threads to release their lock, but note that threads processing queries that map to different buckets cannot interfere with each other.
 The thread then appends the new destination, evicts the least-recently used entry if necessary, and releases the lock again.
Because both critical sections are short, lock contention remains negligible in practice.
We confirm this experimentally in \Cref{sec:experiments}: contention introduces no overhead and has no measurable impact on performance, and end-to-end throughput scales linearly with the number of cores across the evaluated workloads.
\section{Evaluation}
\label{sec:experiments}

We evaluate the performance of \sys and baselines and answer the following questions:
\begin{enumerate}
    \item What is the throughput in terms of queries per second, of \sys and baselines (\diskann and \lshapg) when varying the beam width and number of utilized threads, when using a biased and unbiased query workload (\Cref{sec:exp_biased_workload} and \Cref{sec:exp_unbiased_workload})?
    \item What is the performance of \sys compared to \diskann when using filtered queries (\Cref{sec:exp_filters})?
    \item What is the impact of \sys hyperparameters on achievable throughput (\Cref{sec:exp_hyperparams})?
\end{enumerate}

\subsection{Experimental setup}
\label{sec:exp_setup}

\subsubsection{Workloads}
We evaluate \sys and the baselines using four combinations of query workloads and datasets, \ie, the data used to populate the vector database.

\textbf{\tripclick}~\cite{rekabsaz:2021:tripclick} is a large-scale health-domain search engine log comprising 5.3 million queries issued against a corpus of 35 million \pubmed documents.
Both queries and documents are embedded using \textsc{MedCPT}~\cite{jin2023medcpt} into 768-dimensional vectors, and we populate the vector database with the resulting embeddings of the documents.
Queries are replayed in order, preserving the temporal locality inherent to real user traffic.
This workload with real-world temporal locality serves as our primary benchmark.

\textbf{\medragzipf} shares the same document corpus as \tripclick.
However, we generate a different query workload by prompting a \ac{LLM} to generate semantically equivalent paraphrases of a base set of medical questions, producing clusters of near-duplicate queries in the embedding space.
To construct the query workload, these clusters are sampled according to a Zipfian distribution of parameter 0.8 to simulate the heavy-tailed query frequency skew characteristic of real search logs.
The resulting workload is semi-synthetic since the query distribution is made of embeddings of artificially generated questions.

\textbf{Uniform} consists of \num{150000} query vectors drawn uniformly at random from $[-1, 1]^d$.
We use \textsc{MedCPT} embedding of the first million documents in the \pubmed corpus to populate the vector database.
This fully synthetic query workload is designed to analyze the worst case for \sys: a setting with no spatial or temporal locality.
It allows us to assess whether \sys introduces any throughput regression compared to \diskann when there is no locality to benefit from.

\textbf{Papers} is a natural dataset of \num{10000} queries issued against 2.7 million academic papers uploaded to arXiv.
Documents are represented as 4096-dimensional embeddings extracted from paper abstracts, with each document labeled by its primary arXiv category.
These category labels are used as filter predicates in our filtered search experiments (see \Cref{sec:exp_filters}).

\subsubsection{Baselines}
We compare \sys against two baselines: \diskann and \lshapg.

\textbf{\diskann}~\cite{subramanya:2019:diskann} is a state-of-the-art vector database.
\diskann stores its proximity graph in persistent SSD memory, while compressed vectors (via Product Quantization) are kept in memory for fast approximate distance comparisons during traversal; full vectors are fetched from disk only for the final candidates.
This design scales to billions of vectors on commodity hardware.
We compare the performance of \sys against the base implementation of \diskann.
The search procedure is shown in \Cref{algo:search}, with the medoid of the dataset used as the default starting point.

\textbf{\lshapg}~\cite{zhao:2023:lshapg} augments a proximity graph with a lightweight \ac{LSH}-based index that provides better entry points for graph traversal.
The core idea is that instead of starting every query from the graph medoid, \lshapg first uses hashing to identify a region of the vector space near the query, then launches the graph traversal from a node already close to that region
At index-building time, each indexed vector is mapped to a Z-order hash value, and these values are stored in a sorted numerical index. Binary search on this index quickly retrieves a set of nearby candidates, which are reused as starting points, shortening the search path.

However, \lshapg has practical limitations compared to \sys.
Its \ac{LSH} index must be constructed from scratch alongside the proximity graph, so it cannot be layered on top of an existing index such as \diskann.
The hash functions must also be calibrated to the scale of the dataset upfront, limiting its applicability in streaming scenarios where new vectors arrive over time.
Finally, since the \ac{LSH} structure is built at index time with no awareness of query-time predicates, \lshapg does not support filtered search.
Despite these limitations, \lshapg represents a strong baseline: it is the closest prior work to \sys in spirit. %

\begin{figure*}[t]
	\includegraphics{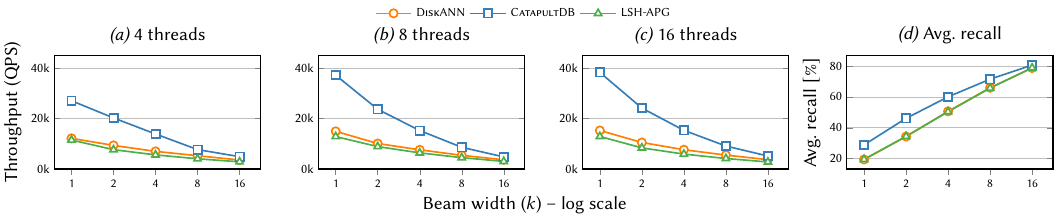}
	\caption{The throughput (in queries per second) and average recall (with four threads, right) of \sys and baselines, while varying the beam width and threads. We use the \medragzipf workload.}
    \label{fig:exp_biased_medrag_qps_k}
\end{figure*}

\begin{figure*}[t]
	\includegraphics{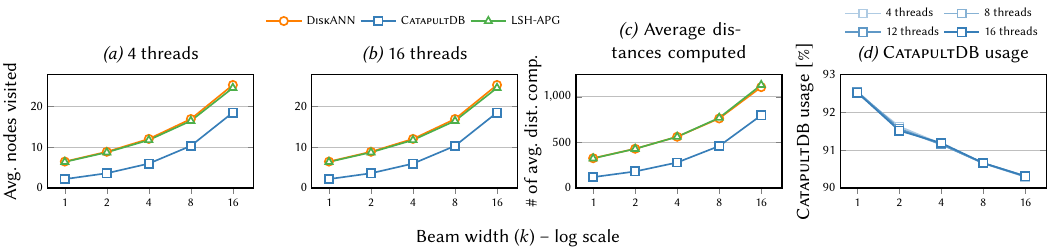}
	\caption{(a,b) The average number of nodes visited in the proximity graph by \sys and baselines, while varying the beam width and threads. (c) The average number of distances computed for \sys and baselines, while varying beam width $k$. (d) The fraction of queries that use a catapult during lookup, for varying beam width and threads. We use the \medragzipf workload.}
    \label{fig:exp_biased_medrag_hitrate}
\end{figure*}

\begin{figure*}
	\includegraphics{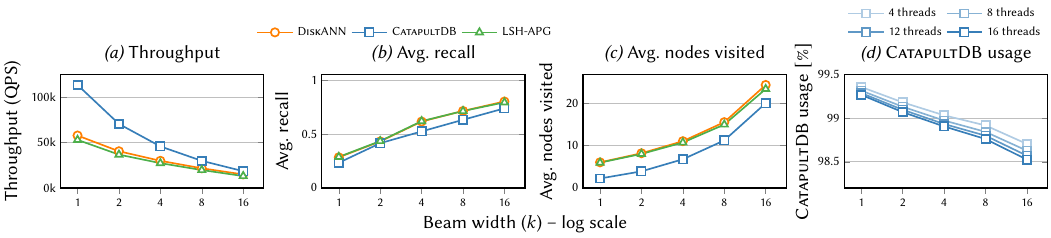}
	\caption{(a) Throughput (in queries per second), (b) average recall with four threads, (c) average nodes visited, and (d) \sys usage, against baseline systems, for different beam width $k$ and number of threads. We use the \tripclick workload.}
    \label{fig:exp_biased_tripclick}
\end{figure*}

\subsubsection{Hardware and implementation}
All experiments were conducted on Docker containers deployed on the EPFL RCP cluster, equipped with Intel Xeon Gold 6240 CPUs and using up to 350\,GB of system memory.
By default, experiments are performed using a unified Rust codebase implementing \diskann-style beam search, \sys, and \lshapg, ensuring a fair comparison across systems.
The experiments with filtered queries enabled (see \Cref{sec:exp_filters}) were conducted on a custom C++ fork of \diskann extended to support the catapult mechanism in a filtering context. Both implementations make use of query-level parallelism and AVX SIMD instructions for distance computations.

\subsubsection{Metrics}
We evaluate system performance along two primary axes: throughput and recall.
Throughput is measured in queries per second (QPS) and serves as our primary performance metric.
We vary two key parameters that directly govern the throughput–recall trade-off.
The \emph{beam width} $k$ controls how many nearest neighbors are retrieved per query and how many candidates are retained during traversal: larger values improve recall at the cost of more computation per query.
The \emph{number of threads} $t$ reflects the degree of query-level parallelism available to the system, which is the dominant scaling mechanism in production deployments where multiple queries are served concurrently.
Recall measures the fraction of true $k$-nearest neighbors returned by the approximate search. We report recall alongside throughput, as throughput improvements are only meaningful when recall is held constant across compared systems.
All experimental results are averaged over three consecutive randomness seeds.

\subsection{Performance of \sys and baselines with a biased workload}
\label{sec:exp_biased_workload}

We now quantify the performance of \sys and baseline systems with the \medragzipf and \tripclick workloads.

\subsubsection{Throughput and recall}
\Cref{fig:exp_biased_medrag_qps_k}(a-c) shows the throughput (in queries per second) of \sys, \diskann and \lshapg, while varying the beam width $k$ and the number of threads $t$, and under the \medragzipf workload.
\sys consistently achieves higher throughput compared to baseline approaches.
For $t=4$ and $k=1$, \sys obtains a throughput of \num{27078.6} QPS, compared to \num{12154.5} and \num{11485.1} QPS for \diskann and \lshapg, respectively.
When increasing $t$, we observe that the throughput of \diskann and \lshapg across values of $k$ remains comparable; \lshapg performs slightly below \diskann because its \ac{LSH} index introduces overhead during traversal.
However, the throughput of \sys increases significantly: from \num{27078.6} QPS for $t=4$ and $k=1$ compared to \num{38333.2} QPS for $k=1$ and $t=16$.
As we increase $k$ and keep $t$ fixed, we also observe that the gains in throughput by \sys compared to baseline systems lower.
For $k=16$ and $t=4$, the throughput of \sys is \num{15328.0}, compared to \num{7623.9} for \diskann, the next-best baseline.
This trend is explained by the nature of the catapult mechanism: finding a good candidate starting point costs a constant amount of work regardless of $k$, while expanding the neighborhood around that candidate scales with $k$.
Catapults accelerate only the former, so their relative contribution to end-to-end query time shrinks as $k$ grows.
\Cref{fig:exp_biased_medrag_qps_k}(d) show the average recall of \sys and baselines.
\sys consistently achieves higher recall and is therefore more effective than \diskann and \lshapg in determining nearest neighbors.
Recall improvement diminishes as $k$ increases: compared to \diskann, \sys increases recall by $48.6\%$ for $k=1$ compared to $2.4\%$ for $k=16$.
Thus, \Cref{fig:exp_biased_medrag_qps_k} shows the efficiency of \sys and achieves up to $2.51\times$ higher QPS than baselines, while also improving or matching recall compared to baselines.

\subsubsection{Compute overhead}
\Cref{fig:exp_biased_medrag_hitrate}(a,b) shows the average number of nodes visited in the proximity graph during a lookup by \sys and baselines, while varying the beam width $k$ and for $t=4$ and $t=16$.
\diskann and \lshapg visit a comparable number of nodes during each lookup.
However, \sys consistently visits fewer nodes in the graph and is therefore able to reach destination nodes quicker.
This reduction is more pronounced for lower values of $k$.
For $ t = 4 $ and $k=1$, \sys visits $66.3\%$ fewer nodes compared to \diskann, compared to $26.9\%$ for $t=4$ and $k=16$.
For all systems, varying $t$ has no significant effect on the number of nodes visited.
\Cref{fig:exp_biased_medrag_hitrate}(c) shows the average distances computed during a lookup for the evaluated systems while varying the beam width $k$.
Compared to \diskann, \sys reduces the number of distance computations by $63.5\%$ and $27.9\%$ for $k=1$ and $k=16$, respectively.
Notably, the reduction in nodes visited and distance computations closely mirror each other across all settings, confirming that \sys's gains stem cleanly from shortcutting graph traversal rather than from any secondary effect, and that the catapult lookup itself introduces negligible overhead.

\subsubsection{Catapult usage}
Finally, we show in \Cref{fig:exp_biased_medrag_hitrate} (right) the percentage of queries that use at least one catapult during lookup for different values of $t$.
\sys obtains high usage, \eg, $92.5\%$ for $k=1$ and $t=4$, highlighting the effectiveness of our approach in speeding up queries.
Usage decreases slightly as $k$ increases, as wider beams are less sensitive to the starting point and thus benefit from catapults on a smaller fraction of their traversal.
Notably, varying $t$ does not affect usage, which is a positive result: despite concurrent writes to shared buckets, the reader-writer locking scheme introduces no measurable staleness or contention, validating the concurrency design of \sys.

\subsubsection{Performance on \tripclick}
\Cref{fig:exp_biased_tripclick} shows the throughput, recall, nodes visited, and catapult usage of \sys and baselines on the \tripclick workload.
The trends closely mirror those observed on \medragzipf.
\sys consistently achieves higher throughput than both \diskann and \lshapg across all values of $k$ and $t$ (up to $89.7\%$), but shows slightly lower recall.
The average number of nodes visited by \sys is up to $62.7\%$ lower than that of the baselines, confirming that catapults effectively reduce the traversal length.
Finally, we observe that catapult usage remains above 98\% across all configurations, demonstrating that \sys leverages the high degree of temporal and spatial locality inherent in the \tripclick query workload.

\begin{figure*}[t]
	\includegraphics{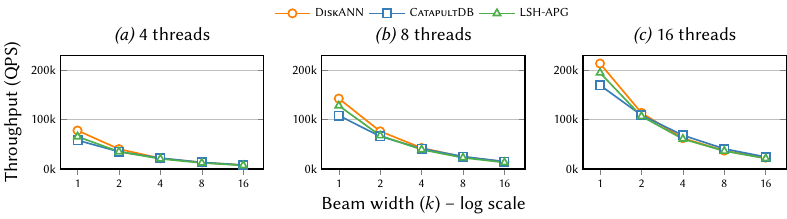}
	\caption{The throughput (QPS) of \sys and baselines, while varying the beam width $k$ and the number of threads. We use the uniform query workload.}
    \label{fig:exp_unbiased_medrag_qps_k}
\end{figure*}

\begin{figure*}[t]
	\includegraphics{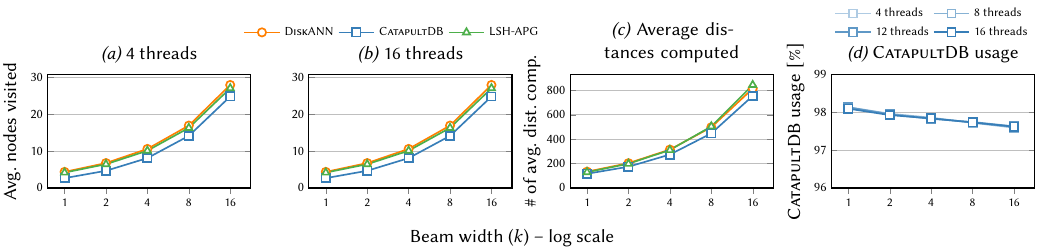}
	\caption{(a,b) The average number of nodes visited in the proximity graph by \sys and baselines, while varying the beam width and threads. (c) The average number of distances computed for \sys and baselines, while varying beam width $k$. (d) The fraction of queries that use a catapult during lookup, for varying beam width and threads. We use the uniform workload.}
    \label{fig:exp_unbiased_medrag_avg_nodes_k}
\end{figure*}

\subsection{Performance of \sys and baselines with an unbiased workload}
\label{sec:exp_unbiased_workload}

We next evaluate the performance of \sys and baselines with the uniform, unbiased workload.
This represents a worst-case scenario where queries are the least likely to benefit from catapults.

\subsubsection{Throughput}
\Cref{fig:exp_unbiased_medrag_qps_k} shows the throughput (in queries per second) of \sys, \diskann and \lshapg, while varying the beam width $k$ and the number of threads $t$.
\sys shows comparable throughput compared to baselines when $k \ge 4$, showing that \sys does not compromise throughput in these settings.
Nevertheless, we notice that there is a performance degradation for $k=1$.
When $t=4$, \sys achieves $25.4\%$ less throughput with $k=1$ compared to \diskann, the best-performing baseline.
As $t$ increases, this decrease becomes less pronounced: for $t=16$, \sys shows $20.6\%$ less throughput with $k=1$ compared to \diskann.
However, we argue that in practice, many \ac{ANN} systems overfetch candidates during traversal to improve recall, and then determine the top-$k$ candidates from the returned list.
Thus, \sys in practice does not compromise on throughput, while providing significant benefits when the query workload contains temporal bias.

\subsubsection{Compute overhead}
\Cref{fig:exp_unbiased_medrag_avg_nodes_k}(a,b) shows the average number of nodes visited in the proximity graph during a lookup by \sys and baselines, while varying the beam width $k$ and for $t=4$ and $t=16$.
Across all settings, \sys reduces the average number of nodes visited, showing that catapults are effective in speeding up query latency, but this decrease is less pronounced than in \Cref{fig:exp_biased_medrag_hitrate}(a,b).
For $t=4$ and $k=1$, \sys decreases the average number of nodes visited by $35.7\%$ compared to the best-performing baseline, which changes to $8.1\%$ for $k=16$.
\Cref{fig:exp_unbiased_medrag_avg_nodes_k}(c) shows the average distances computed during a lookup for the evaluated systems while varying the beam width $k$.
This figure indicates that \sys requires less distance computations than baselines, with the decrease becoming slightly more pronounced as $k$ increases.

\subsubsection{Catapult usage}
\Cref{fig:exp_unbiased_medrag_avg_nodes_k}(d) shows how many queries use a catapult, while varying $k$.
For $k=1$, \sys obtains a usage of $98.1\%$, which decreases to $97.6\%$ for $k=16$.
Remarkably, \sys shows a higher hit rate when using an unbiased query workload compared to a biased workload (see \Cref{fig:exp_biased_medrag_hitrate}(d)).
We attribute this to the effect that uniform queries spread evenly across \ac{LSH} buckets, so catapult destinations accumulate in every bucket and can be reused frequently.
In contrast, the biased workload concentrates traffic in a small number of hot buckets while leaving many other buckets sparsely populated, possibly reducing the overall fraction of queries that find a relevant catapult.
Nevertheless, we note that catapult usage is always above 90\% across all experiments and workloads and conclude that catapults are very frequently used.

\begin{figure}[t]
	\includegraphics{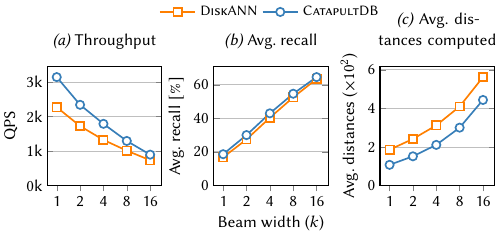}
	\caption{The throughput (in queries per second), average recall and average distance computations for \sys and \diskann, when varying the beam width.}
	\label{fig:exp_filters}
\end{figure}

\subsection{Performance of \sys and \diskann with filter queries}
\label{sec:exp_filters}

We now evaluate the performance of \sys and \diskann in the presence of filter queries, using the \textsc{Papers} workload.
\Cref{fig:exp_filters} shows the throughput (in queries per second), average recall and average distance computations per query for both systems, while varying the beam width.
\Cref{fig:exp_filters}(a) highlights that \sys achieves higher throughput compared to \diskann: $38.47\%$ for $k=1$.
This relative gain in throughput decreases as $k$ increases, \eg, the increase in throughput is $22.10\%$ for $k=16$.
This is a consistent trend across our experiments.
At the same time, \Cref{fig:exp_filters} shows that across values of $k$, \sys achieves slightly higher recall than \diskann, up to $11.01\%$ higher for $k=1$.
\Cref{fig:exp_filters}(c) shows that \sys requires fewer distance computations than \diskann across all values of $k$.
This reduction reflects the shorter traversal paths enabled by catapults: by launching the search closer to the relevant filtered neighborhood, \sys reduces the number of hops and, consequently, the number of distance evaluations required before convergence.
The results in \Cref{fig:exp_filters} confirm that catapults are frequently applicable even under filtered search: the filter predicate disables only a subset of catapult destinations, and the remaining valid destinations are sufficient to provide a meaningful starting-point advantage for the majority of queries.

\begin{figure}[t]
	\includegraphics{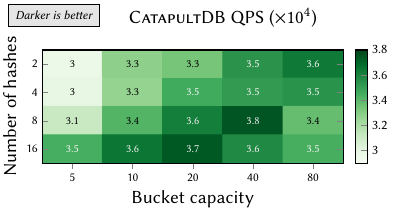}
	\caption{The effect of bucket capacity $b$ and number of \ac{LSH} hyperplanes $L$ on the throughput of \sys, using the \tripclick workload.}
	\label{fig:hyperparams}
\end{figure}

\subsection{The effect of \sys hyperparameters on throughput}
\label{sec:exp_hyperparams}

We evaluate the sensitivity of \sys to its two hyperparameters: the number of \ac{LSH} hyperplanes $L$ and the bucket capacity $b$.
For this experiment, we use the \tripclick workload with $t = 8$ threads and beam width $k = 8$.
\Cref{fig:hyperparams} shows a heatmap of average throughput (QPS) across all $(b, L)$ combinations evaluated.
Across all evaluated configurations, $b = 40$ and $L = 8$ consistently offer the best or near-best throughput, with \num{37552} QPS average across three seeds, representing a 26\% improvement over the weakest configuration ($b = 5$, $L = 2$, \num{29701} QPS).
Therefore, we have selected these values as the default configuration for all other experiments in this paper.
Performance is robust in the vicinity of this optimal point: configurations with $b \in \{20, 40\}$ and $L \in \{8, 16\}$ all achieve throughput similar to the peak, indicating that \sys does not require precise hyperparameter tuning to deliver strong performance.
\section{Related Work}
\label{sec:related}

\paragraph{Graph-based ANN indices.}
Graph-based methods are the dominant paradigm for ANN search~\cite{wang:2021:survey,li:2020:survey}.
They build a proximity graph over the dataset and answer queries via greedy traversal from an entry point.
A central challenge is constructing a graph that is both sparse (low degree per node, to bound the work per hop) and navigable (greedy search converges quickly to the nearest neighbor).

NSW~\cite{malkov:2014:nsw} builds the graph incrementally, inserting vectors one by one and connecting each new vector to its approximate neighbors in the current graph.
This is fast but can produce \emph{hub} nodes with very high degree, which slow down traversal.
HNSW~\cite{malkov:2020:hnsw} addresses this by organizing the graph into multiple levels with a hard cap on the degree at each level, so that no single node becomes a bottleneck; it is arguably the most widely deployed
index in production vector databases~\cite{pan:2024:survey}.
NSG~\cite{fu:2019:nsg} and NSSG~\cite{fu:2022:nssg} pursue sparser graphs with stronger navigability guarantees, while HVS~\cite{lu:2021:hvs} and HCNNG~\cite{munoz:2019:hcnng} reduce construction cost through Voronoi-based and clustering-based strategies, respectively.

\vamana~\cite{subramanya:2019:diskann} introduces a pruning parameter $\alpha > 1$ that explicitly trades a small increase in degree for a much smaller graph diameter, meaning fewer hops per query.
This is the key enabler of \diskann~\cite{subramanya:2019:diskann}, which stores the graph on SSD and keeps only compressed vectors in memory: fewer hops translate directly into fewer disk reads, allowing \diskann to scale to billions of vectors on commodity hardware.
\freshvamana~\cite{singh:2021:freshdiskann} extends \diskann with support for concurrent insertions and deletions while preserving recall over long streams of updates.
\filteredvamana~\cite{gollapudi:2023:filtereddiskann} further adds label-aware graph construction and search, enabling attribute-based filtered queries without sacrificing the efficiency of graph traversal.

In all of these systems, edges are created or removed only when the indexed data changes, never in response to the query workload.
\sys{} is, to the best of our knowledge, the first to reorganize edges on the fly based on observed search trajectories, and is complementary to dynamic and filtered variants such as \freshdiskann and \filteredvamana.

\paragraph{Hybrid LSH--graph approaches.}
\lshapg~\cite{zhao:2023:lshapg} augments a proximity graph with lightweight LSH indexes.
At search time, the LSH structure is queried first to find an entry point that is already close to the query,
shortening the subsequent graph traversal.
It also uses LSH-based distance bounds to prune distant candidates during the traversal itself.
However, \lshapg does not support filtered search, has not been demonstrated at billion-scale, and requires building its own dedicated index: it cannot be layered on top of an existing graph database such as \diskann.
\sys requires no auxiliary structure and derives its shortcuts from the online workload rather than from static hashing at index time.

\paragraph{Learned shortcuts.}
SHG~\cite{gong:2025:shg} trains a piecewise linear model offline to predict how many levels of an HNSW hierarchy can be safely skipped for a given query, avoiding redundant traversal of intermediate levels and
achieving 1.5--1.8$\times$ speedup.
Unlike \sys{}, SHG is specific to hierarchical indices and learns its shortcuts from the data distribution
at construction time, independently of the query workload.
The two approaches are complementary: SHG can shorten the vertical navigation across hierarchy levels, while \sys{} shortcuts the horizontal traversal within a single level.

\paragraph{Caching-based strategies.}
\proximity~\cite{bergman:2025:proximity} places an approximate cache in front of the vector database: if a new query embedding is within a distance threshold~$\tau$ of a previously seen query, the cached neighbor list is returned directly, bypassing the database entirely.
This can eliminate the search cost for highly repetitive queries, but introduces a threshold that must be tuned per workload, does not compose with filtering or dynamic insertions, and requires flushing when the
query distribution shifts.
\sys instead injects shortcut edges into the graph itself, inheriting the full feature set of the underlying index without an external cache.
The two strategies are complementary: \proximity excels when queries repeat almost exactly, while \sys shortens graph traversals.

\section{Conclusion}
\label{sec:conclusion}

We presented \sys{}, a lightweight mechanism that dynamically injects shortcut edges, catapults, into a proximity graph based on observed search trajectories.
Unlike prior graph-based index structures, \sys is the first to reorganize its edge set on the fly to exploit the spatial and temporal locality of the query workload.
Operating as a transparent layer over any index that accepts a configurable entry point, \sys preserves the full feature set of the underlying system, \eg, filtered search, dynamic insertions, and disk-resident indices, without requiring index reconstruction or any modification to the underlying search algorithm.
Our experiments on four workloads demonstrate that \sys as layer on top of \diskann achieves up to $2.51\times$ higher throughput at equivalent or better recall, reducing the average number of nodes visited per query by up to 66.3\% on biased workloads.
\sys also matches or exceeds the search efficiency of \lshapg while additionally supporting filtered queries and out-of-distribution insertions that LSH-APG cannot handle.
Under filtered search, \sys achieves up to 38.47\% higher throughput than \diskann.
In the worst case with a fully uniform, unbiased query workload, \sys introduces no meaningful throughput regression.
\sys therefore represents a step toward workload-aware vector indices, and we anticipate that exploiting query locality can become a standard design consideration in next-generation vector search systems.

\section*{Acknowledgements}
This work has been funded by the Swiss National Science Foundation, under the project “FRIDAY: Frugal, Privacy-Aware and Practical Decentralized Learning”, SNSF grant number 10001796. 

\bibliographystyle{ACM-Reference-Format}
\bibliography{references}

\end{document}